\documentclass[aip,jcp,floats,floatfix,nofootinbib,reprint]{revtex4-1}

\usepackage{amsmath}
\usepackage{amssymb}
\usepackage{amsthm}
\usepackage{multirow}
\usepackage[english]{babel}
\usepackage{graphicx}
\usepackage{xcolor}
\usepackage{lipsum}

\newcommand{\bolda}{{\mathbf{a}}}
\newcommand{\boldu}{{\mathbf{u}}}
\newcommand{\boldx}{{\mathbf{x}}}

\newcommand{\boldzero}{\boldsymbol{0}}
\newcommand{\R}{\mathbb{R}}

\newcommand{\grad}[1]{\ensuremath{\nabla #1}}
\newcommand{\hess}[1]{\ensuremath{\mathcal{H}(#1)}}

\begin{document}

\begin{flushright}
ADP-15-15/T917
\end{flushright}

\title{An index-resolved fixed-point homotopy and potential energy landscapes}

\author{Tianran Chen}
\email{chentia1@msu.edu}
\affiliation{Department of Mathematics, Michigan State University, East Lansing, MI USA.}

\author{Dhagash Mehta}
\email{dmehta@nd.edu}
\affiliation{Department of Applied and Computational Mathematics and Statistics, University of Notre Dame, 
Notre Dame, IN 46556, USA.}
\affiliation{Centre for the Subatomic Structure of Matter, Department of Physics, School of Physical Sciences,
University of Adelaide, Adelaide, South Australia 5005, Australia.}

\begin{abstract}
    \noindent
    Stationary points (SPs) of the potential energy landscapes
    can be classified by their Morse index, i.e., the number of negative eigenvalues of the Hessian evaluated 
    at the SPs.
    In understanding chemical clusters through their potential energy landscapes, 
    only SPs of a particular Morse index are needed.
    We propose a modification of the ``fixed-point homotopy'' method
    which can be used to directly target stationary points of a specified Morse index.
    We demonstrate the effectiveness of our approach by applying it to the 
    Lennard-Jones clusters.
\end{abstract} 
\maketitle

\section{Introduction}

The stationary points (SPs) of the potential energy function, $V(\boldx)$,
with $\boldx = (x_1,\dots,x_n)$, are the bases of the potential energy landscape
methods that have recently gained significant attention from the chemical physics 
community \cite{Wales:04}. Finding SPs of $V(\boldx)$ amounts to solve the corresponding
system of nonlinear equations $\frac{\partial V(\boldx)}{\partial x_i} = 0$, $i = 1,\dots, N$.
The stability and the local geometry of the SPs are determined 
by the eigenvalues of the Hessian matrix of $V(\boldx)$,
$\hess{V}(\boldx)_{ij} =  \partial^2 V/\partial x_i \partial x_j $, evaluated at the SPs. 
While the eigenvalues themselves depend on the choice of coordinate systems,
by the Sylvester's Law of Inertia, their signs are independent from coordinate
systems and hence are intrinsic to the local geometry of the SPs.
The number of negative eigenvalues of $\hess{V}(\boldx)$ is known as
the \emph{Morse index} or simply \emph{index}.
An SP $\boldx$ is said to be \emph{degenerate} if $\hess{V}(\boldx)$ is singular 
and \emph{nondegenerate} otherwise.

A plenty of numerical methods to solve such systems of nonlinear equations 
\cite{morgan1987computing,Nocedal80,lbfgs,TrygubenkoW04,Wales92,Wales93d,munrow99,henkelmanj99,kumedamw01,gwaltney2008interval,
:/content/aip/journal/jcp/140/19/10.1063/1.4875477,DoyeW02,duncan2014biased,crina_trans,Hughes:2014moa,xiao2014basin,henkelman2002methods} 
as well as methods to certify 
numerical approximations \cite{Mehta:2013zia,:/content/aip/journal/jcp/140/22/10.1063/1.4881638} are available
in the energy landscape areas and beyond 
(see Ref.~\onlinecite{mehta2015collection} for a recent short survey).
Among these, homotopy continuation methods have long been used to solve multivariate nonlinear
systems of equations (see Ref.~\onlinecite{allgower_introduction_2003} for surveys in this subject).
Among a rich body of works applying homotopy methods to scientific
problems (e.g. Ref.~\onlinecite{lee_singular_2004}),
recently, the numerical polynomial homotopy continuation method 
\cite{SW05,Mehta:2009,Mehta:2011xs,Mehta:2011wj,Kastner:2011zz,MartinezPedrera:2012rs,Mehta:2013fza,Mehta:2014xya}
has gained special attention for its ability to find all the complex and real SPs of the potentials with polynomial-like
nonlinearities. However, when one is interested in finding only real solutions, this method may turn out to be 
computationally expensive.
The authors have applied another homotopy based method, the Newton homotopy, which 
can be restricted to find only real solutions, with noteworthy results\cite{Mehtacommunication:2014, Mehtafollowup:2015}.
However, this method, same as almost all homotopy based methods, 
does not distinguish among SPs of different indices whereas 
in chemical physics applications, finding stable structures and other 
thermodynamic properties of the systems utilizes only on SPs of indices zero 
(called \emph{minima}) and one (called \emph{transition states}) \cite{Wales:04}.

Among the great variety of homotopy methods, the fixed-point homotopy method
\cite{davidenko_new_1953,scarf_approximation_1967,kellogg_constructive_1976}, 
being one of the first general homotopy methods invented, 
has long been suspected to be able to target index 0. \cite{allgower_homotopy_1979}
Modifications to the fixed-point homotopy method have been proposed 
to target a wider range of SPs, especially index 1 SPs.
Most notable among them was the ``singular fixed-point homotopy'' method
\cite{lee_singular_2004} which uses index 0 SPs as bifurcated starting points
and is likely to be able to reach nearby SPs of index 1.

Several other methods have been developed which attempt to find SPs of 
specific index. Some of the more direct methods include eigenvector following methods \cite{DoyeW02,WalesD03},
single- and double-ended searches and gradient square based methods
\cite{TrygubenkoW04,Wales92,Wales93d,munrow99,%
henkelmanj99,kumedamw01,:/content/aip/journal/jcp/140/19/10.1063/1.4875477,konda2014exploring}, etc.
However, most of these methods are only locally convergent, i.e., one needs to start from an initial guess
close to the actual SP. Moreover, some of these methods also require at least one SP to start with.

In the present contribution, we propose a novel modification to the fixed-point homotopy
which can directly target SPs of a given index.
Unlike the ``singular fixed-point homotopy'' method,
our proposed method starts with random starting points and does not require 
an existing collection of index zero SPs as starting points
(cf.  Ref.~\onlinecite{lee_singular_2004}), and it is hence more likely to 
provide a global sample of index 1 SPs.
This feature is of particular importance when no index zero SP is known.
Compared to other methods based on gradient flow or quasi-Newton techniques,
the proposed method is also likely to have a strong advantage in dealing with
degenerate SPs \cite{Mehtacommunication:2014, Mehtafollowup:2015}.

\section{The deformation of an energy landscape} \label{sec:deformation}

\noindent
In the following, $V(\boldx)$ the potential enrgy function to be explored will be called 
the \emph{target potential energy function}. 
For an integer $m$ with $0 \le m \le n$, called the \emph{target index},
our goal is to locate SPs $\boldx = (x_1,\dots,x_n) \in \R^n$ with index 
exactly $m$. That is, we would like to find points $\boldx \in \R^n$ such that 
\begin{enumerate}
    \item $\grad{V} (\boldx) = \boldzero$; and
    \item $\hess{V}$ at $\boldx$ has $m$ negative eigenvalues.
\end{enumerate}

Criterion 1 amounts to solving a system of nonlinear equations. 
In the present article, we focus on a \emph{homotopy continuation} approach 
for solving this problem\cite{allgower_homotopy_1979,%
allgower_introduction_2003,%
Mehtacommunication:2014,%
Mehtafollowup:2015,li_simple_1979}
in which the target potential function $V$ is embedded into a family of 
closely related potential functions $\hat{V}_t$ parametrized by a parameter 
$t$ so that at $t=1$, $\hat{V}_1 = V$.
Geometrically speaking, this can be understood as the continuous deformation
of the target energy landscape among a family of related landscapes.
The deformation is constructed so that the SPs of one member $\hat{V}_0$ 
at $t=0$ can be studied easily.
The movement of the SPs of $\hat{V}_0$ under this deformation as $t$ varies
from 0 to 1 can then be tracked. In particular, robust numerical algorithms 
can be used to trace the trajectory of the SPs.
\emph{If} the trajectory can be extended to $t=1$, an SP of the 
target potential is then produced as $\hat{V}_1 \equiv V$.

In certain cases, the index can be preserved 
along the trajectory formed by the aforementioned deformation.
By the continuity of eigenvalues of $\hess{\hat{V}_t}$ along a trajectory
it is easy to see that the index must remain the same unless a degenerate SP 
of $\hat{V}_t$ is encountered where at least one eigenvalue of
$\hess{\hat{V}_t}$ vanishes.\footnote{
This is proved in Ref.~\onlinecite{allgower_homotopy_1979} and used without explicit 
statement in several related works e.g. 
Ref.~\onlinecite{davidenko_new_1953,abraham_transversal_1967}.}
However, this criterion is quite limited, since from the point of view of
\emph{catastrophe theory}\cite{poston_catastrophe_2014,wales2001microscopic} 
as one traces the SPs of a family of potential functions, an encounter with 
a degenerate SP is not only possible, but sometimes necessary.

In the following we propose a new homotopy construction that
can potentially target SPs of a given index under much relaxed conditions.

\section{Index-resolved fixed-point homotopy} \label{sec:morse-homotopy}

The fixed-point homotopy \cite{davidenko_new_1953,scarf_approximation_1967} 
is one of the first homotopy methods devised to solve general nonlinear systems.
Here we propose an ``\emph{index-resolved fixed-point homotopy}''
method for finding SPs of a given index.
As we shall demonstrate in \S \ref{sec:lj-cluster} with experiments on the
Lennard-Jones cluster, it is capable of quickly obtaining a large number of
SPs of a given index.

In keeping with the general framework for finding SPs of a smooth 
potential using homotopy methods described above,
we construct a family of closely related potential $\hat{V}_t$
parametrized by $t$ so that $\hat{V}_1 \equiv V$ yet the SPs of the 
\emph{starting potential} $\hat{V}_0$ can be located and studied easily.
Here the starting potential is chosen to be the ``standard $m$-saddle''
located at a point $\bolda = (a_1,\ldots,a_n)$ which is represented by the potential
\begin{scriptsize}
    \begin{equation}
        \label{equ:saddle1}
        -(x_1-a_1)^2 - \cdots -(x_m-a_m)^2 + (x_{m+1}-a_{m+1})^2 + \cdots + (x_n-a_n)^2.
    \end{equation}
\end{scriptsize}%
Clearly, $\boldx = \bolda$ is a nondegenerate SP of the above function with 
index exactly $m$. For brevity, we define the quadratic form as
\begin{equation}
    \label{equ:J}
    J^{(m)}(\boldu) =  
    \frac{1}{2}
    \begin{bmatrix}
        u_1 & \cdots & u_n
    \end{bmatrix}
    \begin{bmatrix}
        -I_m &         \\
             & I_{n-m}
    \end{bmatrix}
    \begin{bmatrix}
        u_1 \\ \vdots \\ u_n
    \end{bmatrix},
\end{equation}
where $\boldu=(u_1,\ldots,u_n)$, and $I_m$ and $I_{n-m}$ are identity 
matrices of dimension $m \times m$ and $(n-m) \times (n-m)$, respectively.
The standard $m$-saddle \eqref{equ:saddle1} at $\boldx = \bolda$ 
is then represented by the quadratic function $J^{(m)}(\boldx-\bolda)$.

Representing the deformation between this $m$-saddle and the 
target potential $V$ is the one-parameter family
\begin{equation}
    \label{equ:morse-deformation}
    \hat{V}^{(m)}_t(\boldx) := (1-t)\, J^{(m)}(\boldx-\bolda) + t\, V(\boldx).
\end{equation}
This family contains the target potential $V(\boldx)$ as a member, 
since at $t=1$, $\hat{V}^{(m)}_1 \equiv V$. 
The starting potential $\hat{V}^{(m)}_0(\boldx) \equiv J^{(m)}(\boldx-\bolda)$,
at $t=0$, has a unique SP of index $m$. 
As $t$ varies, $\hat{V}^{(m)}_t$ represents a smooth deformation 
between the two potentials.

We now consider the effect of the deformation \eqref{equ:morse-deformation} on the 
SP $\boldx = \bolda$, that is, how the SP $\boldx = \bolda$ of $\hat{V}^{(m)}_0$ 
relates to SPs of nearby members of $\hat{V}^{(m)}_t$ as $t$ moves away from $t=0$.
The local theory of smooth real-valued functions yields that under a small 
perturbation, nondegenerate SPs remain nondegenerate, i.e., by varying $t$ 
sufficiently close to 0, the nondegenerate SPs of $\hat{V}^{(m)}_t$ traces out a 
small piece of smooth curve.
The \emph{homotopy continuation} approach hinges on the ``continuation'' of this
small piece of curve into a globally defined curve that could potentially 
connect the SP $\boldx=\bolda$ of $\hat{V}^{(m)}_0$ to an SP of the target potential 
$\hat{V}^{(m)}_1 \equiv V$ which we aim to find.
A standard application of the \emph{Generalized Sard's Theorem}
\cite{abraham_transversal_1967} guarantees that this is possible for ``generic''
choices of $\bolda$.
Putting aside the technical statement of the theorem, it essentially means that
if $\bolda$, the (unique) SP of the starting potential, is chosen at random,
then with \emph{probability one}\footnote{Equivalently, this is stating that
the probability of choosing the ``bad'' start points is zero.}
the set of all SPs of $\hat{V}^{(m)}_t$ for $t \in [0,1)$ form of smooth curves
in the $(n+1)$-dimensional $\boldx$-$t$ space.
Among them, we focus on the curve that passes through $\boldx = \bolda$ and $t=0$,
which will be called the \emph{trajectory} of $\bolda$.
Since $\hat{V}^{(m)}_1 \equiv V$, if this trajectory extends to $t=1$ then an SP
of the target potential is obtained.

From the algebraic standpoint, if we define the 
\emph{index-resolved fixed-point homotopy with index $m$} as
\begin{equation}
    \label{equ:morse-homotopy}
    H^{(m)}(\boldx,t) := \grad{\hat{V}}^m_t(\boldx),
\end{equation}
then the crux of the method is to numerically trace along the smooth trajectory
which is a curve defined by $H^{(m)}(\boldx,t) = \boldzero$ and containing the point
$(\bolda,0)$ in the hope that a point $(\boldx^{(1)},1)$ can be reached
\footnote{Though not limited to gradient systems, the original fixed-point homotopy 
    \cite{davidenko_new_1953,scarf_approximation_1967,kellogg_constructive_1976}
    can be considered as a special case of this construction with the target index 0, 
    i.e., the homotopy $H^{(0)}$.}.

While there is a rich selection of numerical methods for tracing such curves, 
within the community of homotopy methods, a special class of methods based on 
a \emph{predictor-corrector} scheme together with an \emph{arc-length parametrization}
of the curve has emerged as the method of choice for its superior 
stability and efficiency 
\cite{Mehtacommunication:2014,allgower_introduction_2003,chu_numerical_1983,allgower_survey_1981,li_simple_1979,chu_simple_1984,SW05}.

Our goal, however, is to locate SPs of index $m$.
Since the starting point $\boldx = \bolda$ is an index-$m$ SP of the starting
potential $\hat{V}^{(m)}_0$, the hope is that this index is inherited by the
SP of the target potential the trajectory obtains (if the trajectory reaches $t=1$).
Clearly, if the trajectory encounters no degenerate SP, the continuity of
eigenvalues of $\hess{\hat{V}^{(m)}_t}$ (which must remain nonzero without 
encountering a degenerate SP) ensures preservation of index along the entire 
trajectory, and consequently the SP of the target potential it reaches
has index $m$.
However, our numerical experiments with the Lennard-Jones potential 
(\S \ref{sec:lj-cluster}) suggest that the index can be preserved 
even when the assumption in this theorem fails to hold, 
that is, when the trajectory encounters some turning points.
Some observations are presented in \S \ref{sec:index}.

\subsection{Summary of the Proposed Numerical Algorithm} \label{sec:algo}

For the target potential function $V(\boldx)$ and a given target index 
$m \le n$, one first randomly pick a starting point $\bolda \in \R^n$ 
with which the family
\[
    \hat{V}^{(m)}_t (\boldx) := (1-t) \, J_m (\boldx - \bolda) + t \, V(\boldx)
\]
is constructed. 
We are interested in the SPs of $\hat{V}^{(m)}_t$ for all $t$ as a set in the 
$(n+1)$-dimensional space. 
This set is defined by the nonlinear system 
$H^{(m)}(\boldx,t) := \grad{\hat{V}^{(m)}_t} = \boldzero$.
The starting point $(\bolda,0)$ is clearly in this set.
Moreover, for generic choices of $\bolda$, this system defines smooth trajectory
with the starting point $\boldx = \bolda$ and $t=0$.

One can then apply efficient numerical methods based on the predictor-corrector 
scheme to trace along the smooth trajectory.
\emph{If} the trajectory extends to $t=1$, i.e., if it passes through a point 
$(\boldx,t) = (\boldx^{(1)},1)$, then an SP of the target potential function $V$ 
is produced\footnote{To control the ``numerical condition'' of the curve tracing algorithm,
    this formulation of the deformation $\hat{V}^{(m)}_t$ can be generalized to
    \[
        \tilde{V}^{(m)}_t (\boldx) := 
        (1-t) \, J^{(m)} (P(t) \cdot (\boldx - \bolda)) + t \, V( Q(t) \cdot \boldx)
    \]
    where for each $t$ value $P(t)$ and $Q(t)$ are nonsingular $n \times n$ matrices 
    which can be chosen to tame the numerical condition of the corresponding 
    system of equations $\grad{\tilde{V}^{(m)}_t(\boldx)} = \boldzero$.
    For example, by choosing $P(t)$ and $Q(t)$ to be diagonal matrices
    with positive entries, one can selectively scale the individual variables
    $x_1,\dots,x_n$ to ease the potential problems of imbalance in magnitudes.}.

\subsection{An example} \label{sec:example}

\begin{figure}[h]
    \includegraphics[width=0.75\columnwidth]{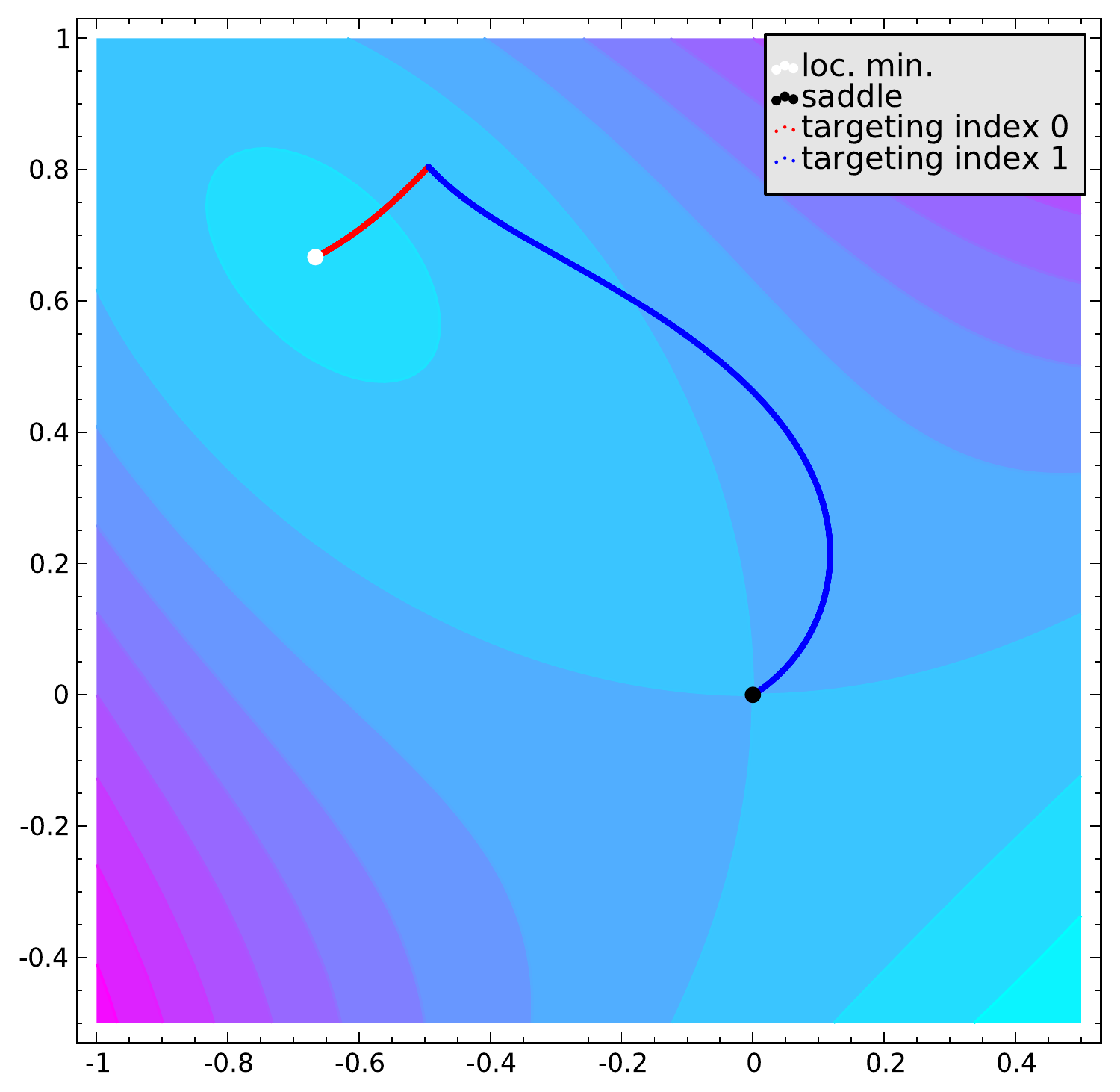}
    \caption{
        Contour plot of potential function \eqref{equ:example}
        which has a local minimum (index 0) at $(-2/3,2/3)$
        and a saddle point (index 1) at $(0,0)$.
        Starting from the exact same point the two index-resolved fixed-point 
        homotopies with different target indices $\hat{V}^0_t$ and $\hat{V}^1_t$ 
        define two different trajectories marked by the red/light 
        and the blue/dark curves respectively.
        }
    \label{fig:example-contour}
\end{figure}

\noindent
As an example, consider the potential function
\begin{equation}
    \label{equ:example}
	V(x,y) = -x^3 + y^3 + 2xy - 6,
\end{equation}
whose contour plot is shown in Figure \ref{fig:example-contour}. It has two SPs: 
a local minimum (index 0) at $(-2/3,2/3)$ and a saddle point (index 1) at $(0,0)$.
\vskip 1ex

\noindent \textbf{(Index 0)}
To locate the index 0 SP we construct
\begin{equation}
    \label{equ:example-V0}
    \hat{V}^{(0)}_t := \frac{1-t}{2} [(x-a_1)^2 + (y-a_2)^2] + t \, V(x,y)
\end{equation}
according to \eqref{equ:morse-deformation} with target index $m=0$ where 
$(a_1,a_2) = (-0.494332,0.804689)$ is a randomly chosen starting point.
Clearly, $(x,y) = (a_1,a_2)$ is an SP of index 0 of the 
starting potential $\hat{V}^{(0)}_0 = (x-a_1)^2 + (y-a_2)^2$.

The projection of the smooth trajectory, defined by 
$H^{(0)} (x,y) := \grad{\hat{V}}_t^{(0)}(x,y) = (0,0)$, onto the $x$-$y$ plane 
(where $t$ is removed) is shown in Figure \ref{fig:example-contour} as the 
red/light curve. It converges to the index 0 SP at $(-2/3,2/3)$.
\vskip 1ex

\noindent \textbf{(Index 1)}
To locate the index 1 SP, we construct
\begin{equation}
    \label{equ:example-V1}
    \hat{V}^{(1)}_t := \frac{1-t}{2} [-(x-a_1)^2 + (y-a_2)^2] + t V(x,y)
\end{equation}
with the only difference from \eqref{equ:example-V0} being the sign of $(x-a_1)^2$.
With the same starting point $(a_1,a_2) = (-0.494332,0.804689)$, 
$H^{(1)}(x,y) := \grad{\hat{V}}_t^{(1)}(x,y) = (0,0)$ defines a 
different trajectory whose projection onto the $x$-$y$-plane 
is shown in Figure \ref{fig:example-contour} as the blue/dark curve. 
This curve converges to the saddle point $(0,0)$ of $V(x,y)$ instead.
It is worth pointing out that even though the starting point $(a_1,a_2)$ is
very close to the index 0 SP, as shown in Figure \ref{fig:example-contour}, 
the trajectory still converges to the index 1 SP which is much further way.

In both cases, the target indices $m$ in \eqref{equ:example-V0}
and \eqref{equ:example-V1} determine which SP the corresponding curve converge to.

\section{The Lennard-Jones Cluster}\label{sec:lj-cluster}
We apply our method to find SPs of the Lennard-Jones cluster
\cite{jonesi25} of $N$ atoms whose potential function is given as
\begin{equation}
    V_N = 4\epsilon \sum_{i=1}^N \sum_{j=i+1}^N 
	\left[ 
	    \left( \frac{\sigma}{r_{ij}} \right)^{12} - \;
	    \left( \frac{\sigma}{r_{ij}} \right)^{6} 
	\right],
    \label{equ:lj-cluster}
\end{equation}
where $N$ is the number of atoms, $2^{1/6}\sigma$ is the equilibrium pair separation, $\epsilon$ is the pair well depth, and
\[ r_{ij} = \sqrt{(x_i - x_j)^2 + (y_i - y_j)^2 + (z_i-z_j)^2} \]
is the Euclidean distance between an $i$-th and $j$-th atoms.
We take $\epsilon = \sigma = 1$ for simplicity. In addition to the fact that 
the Lennard-Jones cluster potential serves as a good approximate for the atomic interactions,
the potential exhibits complicated landscape structure, i.e., the number of minima exponentially 
grows when increasing $N$ and has a multi-funnel structure due to the simultaneous
presence of competing growth sequences \cite{Wales:04}. This model has been under extensive searches for minima and higher index SPs 
\cite{DoyeW02,2005JChPh.122h4105D} which provided us bases for comparison.

Defined in terms of the pairwise distances,
$V_N$ is clearly invariant under rotation and translation.
Consequently in the $x$-$y$-$z$ coordinates, all SPs would have certain degrees
of freedom. After a translation of the cluster, we can fix the first atom at 
the origin, i.e., $x_1 = y_1 = z_1 = 0$.
For the ease of computation, we shall restrict our attention to the cases 
where the $N$ atoms are not collinear. We can therefore fix
$y_2 = z_2 = z_3 = 0$ and also require $y_3 \ne 0$.\footnote{
    If the $N$ atoms are not collinear, after relabeling the $N$ atoms
    and rotations, we can always find a representative that satisfy these conditions.
    For the pathological collinear configurations (where all $N$ atoms lie on a line)
    the restriction $y_2 = z_2 = z_3 = 0$ does not remove all the degrees of freedom.
    Consequently such configurations are best handled by alternative formulations.
}
After the restriction, we have a total of $3N-6$ variables in $V_N$. 
Due to the permutation symmetry, $V_N$ may exhibit the same value at different SPs. 
Such SPs are known as \emph{permutation-inversion isomers}. \cite{wales2010energy,calvo2012energy}

\begin{table*}[t]
    \begin{tabular}{|c|c||c|c|c|c|c|c|c|c|c|c|c|}
    	\hline
    	Target & Obtained &                                             \multicolumn{11}{c|}{$N$}                                              \\ \cline{3-13}
    	index  &  index   &   5    &    6    &    7    &    8     &    9     &    10    &    11    &    12    &    13    &    14    &    15    \\ \hline\hline
    	$m=0$  &    0     & 1 (22) & 2 (59)  & 4 (101) & 8  (134) & 15 (101) & 28 (132) & 45 (46)  & 58 (193) & 68 (110) & 60 (92)  & 70 (78)  \\ \hline
    	$m=1$  &    1     & 2 (53) & 3 (92)  & 8 (105) & 13 (148) & 22 (100) & 57 (391) & 65 (243) & 80 (385) & 79 (313) & 70 (96)  & 64 (242) \\ \hline
    	$m=2$  &    2     & 4 (44) & 6 (168) & 8 (142) & 14 (170) & 14 (96)  & 15 (106) & 56 (124) & 52 (143) & 57 (119) & 55 (115) & 52 (97)  \\ \hline
    \end{tabular}
    \caption{
        Numbers of geometric configurations of $N$ atoms 
        (together with numbers of SPs representing them) with different indices 
        of the Lennard-Jones potential $V_N$ \eqref{equ:lj-cluster} for a range 
        of $N$ values obtained by index-resolved fixed-point homotopies
        with target indices $m=0,1,2$ respectively.
        E.g., the entry ``2 (59)'' for $N=6$, $m=0$, and ``obtained index'' 0
        indicates that a total of 59 SPs of $V_6$ having index 0 are found using
        the homotopy $H^{(0)}$, and under the permutation symmetry they collectively 
        represent two geometric configurations of the 6 atoms.
        \emph{No nondegenerate SPs of any other indices were found in these runs.}
    }
    \label{tab:lj}
\end{table*}

We applied the index-resolved fixed-point homotopy to the problem of finding SPs 
of the Lennard-Jones potential $V_N$ with certain indices.
Since SPs of indices 0 (local minima), 1 (transition states), and 2 are most 
frequently used in computational chemistry, we restrict our attention
to homotopies $H^{(m)}$ (as defined in \eqref{equ:morse-homotopy}) with $m=0,1,2$,
although general constructions with any $m \le 3N-6$ are possible.
\cite{DoyeW02,2005JChPh.122h4105D}
Table \ref{tab:lj} shows the number of SPs of $V_N$ for a range of $N$ values 
obtained by $H^{(0)}$, $H^{(1)}$, and $H^{(2)}$ respectively.
Remarkably, the indices of \emph{all} SPs found match exactly the target index $m$ 
used in the construction of $H^{(m)}$, suggesting that the index-resolved fixed-point 
homotopy proposed here has a great potential in targeting SPs of a specific index.

As a homotopy-based method, the index-resolved fixed-point homotopy has a 
strong advantage in dealing with degenerate SPs.\cite{Mehtacommunication:2014}
The runs that produced SPs listed in Table \ref{tab:lj} have also resulted in
a number of SPs that appear to be degenerate (and hence not counted).
Table \ref{tab:lj-degenerate} lists a selection of these 
\emph{numerically degenerate} SPs where Hessian matrices $\hess{V_N}$ 
have eigenvalues of magnitude less than $10^{-10}$.
Though the physical meaning of the degenerate SPs in the Lennard-Jones clusters
appear to be understudied,
their very presence highlights the richness of the energy landscape
and merits further analysis.

\begin{table}[h]
    \begin{tabular}{|c|c|c|c|c|}
    	\hline
    	$N$ &      $V_N$ at SP      & \# neg. eig. & \# zero eig. & Min. eig.  \\ \hline
    	 5  & $-3.0000000002858558$ &      0       &      6       & $10^{-13}$ \\ \hline
    	 5  & $-4.0000000000004334$ &      0       &      5       & $10^{-19}$ \\ \hline
    	 5  & $-6.0000000000157154$ &      0       &      3       & $10^{-14}$ \\ \hline
    	 6  & $-6.0000000007539667$ &      0       &      6       & $10^{-13}$ \\ \hline
    	 6  & $-8.4806201568350339$ &      1       &      3       & $10^{-12}$ \\ \hline
    	 6  & $-9.1038524157210112$ &      0       &      3       & $10^{-14}$ \\ \hline
         7  & $-9.1038524162043597$ &      0       &      6       & $10^{-13}$ \\ \hline
    	 8  & $-12.302927529932436$ &      0       &      6       & $10^{-13}$ \\ \hline
    	 8  & $-14.811129645604781$ &      1       &      3       & $10^{-13}$ \\ \hline
    	 8  & $-15.444733800924601$ &      1       &      3       & $10^{-12}$ \\ \hline
    	 8  & $-15.533060054591045$ &      0       &      3       & $10^{-15}$ \\ \hline
    	 8  & $-16.505384168030663$ &      0       &      3       & $10^{-15}$ \\ \hline
    	 9  & $-18.597348435556142$ &      1       &      3       & $10^{-12}$ \\ \hline
    	 9  & $-18.778208174125020$ &      0       &      3       & $10^{-14}$ \\ \hline
    	 10 & $-18.856826168132727$ &      0       &      6       & $10^{-16}$ \\ \hline
    \end{tabular}
    \caption{
        A selection of \emph{numerically degenerate} SPs of $V_N$ obtained by the
        index-resolved fixed-point homotopy \eqref{equ:morse-homotopy} for a range
        of $N$ values.
        For each SP, ``$V_N$ at the SP'' is the value of $V_N$ evaluated at the SP.
        ``\# neg. eig.'' is the number of negative eigenvalues of $\hess{V_N}$ at
        the SP.
        ``\# zero eig.'' is the number of eigenvalues of $\hess{V_N}$ at the SP 
        that have magnitude less than $10^{-10}$ and \emph{do not} correspond to 
        the inherent degree of freedom induced by rotation and translation.
        ``Min. eig.'' shows the approximate scale of the of the smallest magnitude
        of the eigenvalues counted in the column ``\# zero eig.''.
    }
    \label{tab:lj-degenerate}
\end{table}

\section{Preservation of index under relaxed conditions} \label{sec:index}

\begin{figure}[h]
    \includegraphics[width=0.85\columnwidth]{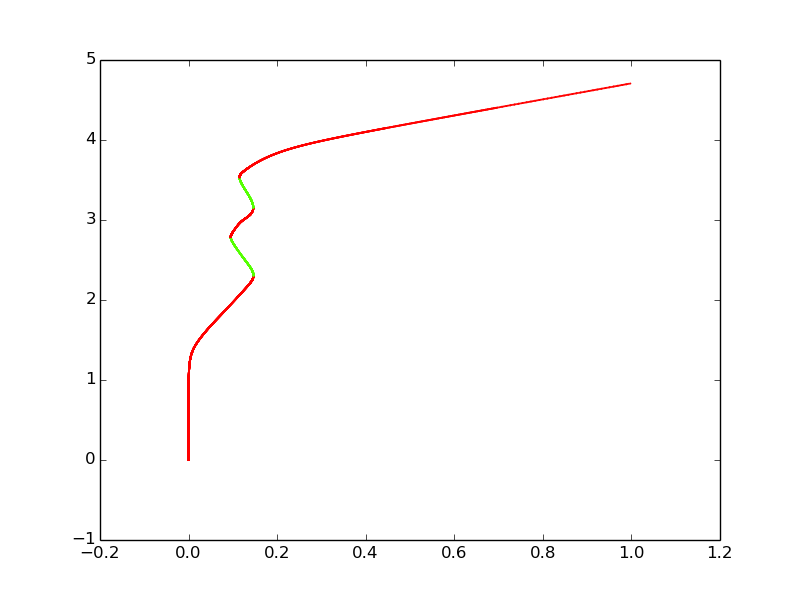}
    \caption{
        The $t$ value (horizontal) plotted against the arc-length (vertical) 
        along a trajectory of the index-resolved fixed-point homotopy with $N=7$ and $m=2$.
        For each point $(\boldx,t)$, the index of $\boldx$ as an SP of $\hat{V}^{(2)}_t$ 
        is indicated by color: 
        red/dark for index 2 and green/light for index 1.
    }
    \label{fig:lj-index}
\end{figure}

As discussed in \S \ref{sec:deformation}, by the continuity of eigenvalues of 
$\hess{\hat{V}^{(m)}_t}$ along the trajectory of $\boldx = \bolda$,
the index must remain the same as long as no degenerate SP of $\hat{V}^{(m)}_t$ 
is encountered (where $\hess{\hat{V}^{(m)}_t}$ becomes singular).
However, our numerical experiments with the Lennard-Jones cluster (\S\ref{sec:lj-cluster})
suggest that the index can be preserved under much more relaxed conditions.
In particular, we observed that \emph{index can be preserved even when 
turning points are encountered.}
Figure \ref{fig:lj-index} shows the $t$-value plot together with the changes of
indices along a curve defined by $H^{(2)}=\boldzero$ for the Lennard-Jones cluster potential 
\eqref{equ:lj-cluster} with $N=7$ and target index $m=2$. 
After four changes (from index 2 to 1, then 2, 1, and back to 2),
the index comes back to the target index $m=2$.

Among our experiments summarized in Table \ref{tab:lj}, such phenomenon
where trajectories encounter some degenerate SPs before reaching SPs 
of the target potential appear to be quite common.
Indeed, we estimate that at least $98\%$ of the SPs counted in Table \ref{tab:lj}
for $N=10,\dots,15$ 
(where all SPs obtained have indices agreeing with target indices) 
were obtained by curves that encounter degenerate SPs, suggesting that a much weaker condition may exist for the 
preservation of indices.

\section{Conclusion}

Specializing for the potential energy landscape scenarios, 
we have proposed a novel homotopy continuation based method, 
called index-resolved fixed-point homotopy, which can find SPs of specific index. 
It does not require a preexisting set of SPs as its starting points. 
The method can also find certain singular SPs of particular index without much
numerical difficulties unlike other Newton or quasi-Newton based methods. 
With our numerical experiments with the Lennard-Jones clusters, we have also demonstrated that
the proposed homotopy continuation approach can target SPs of a specific 
index with mild conditions on the potentials.
This result may trigger further activities in such specialized homotopy 
continuation based approaches among the relevant mathematics community. 
Future work will involves the rigorous analysis of this homotopy method 
as well as a systematic comparison of our proposed approach with various 
existing methods to find the SPs of a specific index.

\begin{acknowledgments} 
DM was supported by a DARPA Young Faculty Award 
and an Australian Research Council DECRA fellowship.
TC was supported in part by NSF under Grant DMS 11-15587.
TC would like to thank the Institute for Cyber-Enabled Research at 
Michigan State University for providing the computational infrastructure. 
We thank Jonathan Hauenstein, Cheng Shang and David Wales for their feedback on this work. 
\end{acknowledgments}

\bibliography{morse,bibliography_NPHC_NAG,bibliography_NPHC_NAG_Crina,invert}

\end{document}